\documentclass[conference,a4paper]{IEEEtran}
\usepackage[dvips]{graphicx}
\usepackage{amsmath,amssymb}

\newcommand{\defeq}{\stackrel{\triangle}{=}}

\newtheorem{theorem}{Theorem}

\newcommand{\qed}{\fbox{}}
\newif\ifhoge
\hogetrue

\begin{document}
\sloppy

\title{Finite Length Analysis on Listing Failure Probability of Invertible Bloom Lookup Tables}

 \author{
   \IEEEauthorblockN{
     Daichi Yugawa\IEEEauthorrefmark{1} and
     Tadashi Wadayama\IEEEauthorrefmark{1}}
   \IEEEauthorblockA{
     \IEEEauthorrefmark{1}Department of Computer Science and Engineering, \\
     Nagoya Institute of Technology Nagoya, Japan\\ Email: yugawa@cs.it.nitech.ac.jp, wadayama@nitech.ac.jp}
 }



\maketitle

\begin{abstract}
The Invertible Bloom Lookup Tables (IBLT) is a data structure which supports 
insertion, deletion, retrieval and listing operations of the key-value pair. The IBLT can be used 
to realize efficient set reconciliation for database synchronization. The most notable feature of 
the IBLT is the complete listing operation of the key-value pairs
based on the algorithm similar to the peeling algorithm for low-density generator-matrix (LDGM) codes.
In this paper, we will present a stopping set (SS) analysis for the IBLT which reveals 
finite length behaviors of the listing failure probability.   The key of the analysis is enumeration of the 
number of stopping matrices of given size. We derived a novel recursive formula useful for computationally efficient enumeration.
An upper bound on the listing failure probability based on the union bound accurately captures the error floor behaviors.
It will be shown that, in the error floor region, the dominant SS have size 2. We propose a simple modification on 
hash functions, which are called SS avoiding hash functions,  for preventing occurrences of the SS of size 2.
\end{abstract}

\section{Introduction}

The Invertible Bloom Lookup Tables (IBLT) is a recently developed data structure which 
supports insertion, deletion, retrieval and listing operations of the key-value pairs \cite{IBLT}--\cite{BC}.
The IBLT can be seen as a natural extension of the Bloom filter \cite{BF}--\cite{CBF} which can handle set membership queries. The most notable feature of the IBLT is the complete listing operation of the key-value pairs
based on the algorithm similar to the peeling algorithm \cite{richardson} for low-density generator-matrix (LDGM) codes.

The listing operation enable us to use the IBLT for a basis of an efficient set reconciliation algorithm with small amount of communications.
Set reconciliation is a process to synchronize contents of two sets at two distinct locations and
it can be used for realizing database synchronization, 
memory synchronization, and an implementation of the Biff codes \cite{BC}. The implementation of the IBLT is fairly simple and it is 
naturally scalable to multiple servers, which is a desirable feature for data sets of extremely large size.

The paper by Goodrich and Mitzenmacher \cite{IBLT} provides the detailed analysis on the IBLT such as 
the optimization of the number of hash functions to minimize the retrieval failure probability.
They also presented asymptotic thresholds for accurate recovery by using the known results on 2-cores of random hypergraphs.
Furthermore, some fault tolerance features of the IBLT are extensively studied.

For designing practical applications, it is beneficial to know not only the asymptotic behavior of listing processes 
but also finite length performances. Especially, predicting the error floor of the listing failure probability is required to 
guarantee the accuracy of a listing process. It is known {\em stopping sets} \cite{di} dominate the finite length performance of
LDGM codes  for erasure channels. In the case of the IBLT, the stopping sets have crucial importance as well as the case of 
LDGM codes.
In this paper, we will present a stopping set analysis for the IBLT which unveils
the finite length behaviors of the listing failure probability.  

The outline of this manuscript is organized as follows.
Section \ref{pre} introduces notation and definitions required for this paper. 
A brief review of the IBLT is also given.
Section \ref{upper} provides an upper bound on the listing failure probability.
An enumeration method for the number of stopping matrices based on a recursive formula
is the heart of the efficient evaluation of the upper bound.
Section \ref{numerical} presents some results of computer experiments.
It will be shown that, in the error floor region,  the stopping sets with size 2 become dominant.
In Section \ref{lowering}, a class of hash functions, SS avoiding hash functions, is proposed to resolve the
stopping sets with size 2 for lowering the error floor.

\section{Preliminaries}
\label{pre}

\subsection{Bloom Filter}
Before going into details of the IBLT, we here explain the structure of the original Bloom filter (BF) which is the basis of the IBLT.
Assume that we have a binary array $T$ and $k$-hash functions $h_1,\ldots, h_k$.
The binary array $T$ is initially set to all zero. When an item $x$ comes to insert,
we set $T[h_i(x)]=1$ for $i \in [1,k]$. The notation  $[\alpha, \beta]$ means the set of consecutive integers from $\alpha$ to $\beta$. The process is called the {\sf Insert}$(x)$ operation.
The set membership query on $y$ is the query for checking whether $y$ is in the BF or not.
The {\sf LookUp}$(y)$ operation returns YES if $T[h_i(y)]=1$ for $i \in [1,k]$; otherwise it returns NO.
The operations  {\sf Insert}$(x)$ and {\sf LookUp}$(y)$ can be carried out in $O(k)$-time.
Note that the {\sf LookUp}$(y)$ operation may yield false positive; i.e., it returns YES when $y$ is not in the BF.
The minimization of this false positive probability in terms of the number of hash functions is an important topic 
of studies of the BF \cite{BF}\cite{CBF}. An appropriately designed BF provides a highly space efficient 
set membership query system with reasonably small false positive probability.

\subsection{IBLT and its Operations}

As in the case of the BF, $k$-hash functions $h_1,\ldots, h_k$ are used in the IBLT.
Instead of binary array, the IBLT utilizes an array of {\em cells} $T[1],\ldots, T[m]$.
A cell $T[i]$ consists of three fields which are called {\em Count, KeySum,} and {\em ValueSum}, which 
are denoted by $T[i].Count, T[i].KeySum, T[i].ValueSum$.
An input to the IBLT is a key-value pair $(Key, Value)$. The count field represents the number of
inserted entries. The KeySum (resp. ValueSum) field stores exclusive OR of key (resp. value) of inserted entries.
The contents of all the cells are initialized to zero at the beginning.

The IBLT allows 4-operations: {\sf Insert}$(x,y)$, {\sf Delete}$(x,y)$,  {\sf Get}$(x)$ and {\sf ListEntries}$()$.
The operation {\sf Insert}$(x,y)$ stores a key-value pair $(x,y)$ into the IBLT.
In an insertion process,  the key $x$ (resp. value $y$) is added (over $\Bbb F_2$) to the KeySum (resp.  ValueSum) filed 
of $T[h_i(x)]$ for $i \in [1,k]$; namely, $T[h_i(x)].KeySum=T[h_i(x)].KeySum\oplus x$ and $T[h_i(x)].ValueSum=T[h_i(x)].ValueSum\oplus y$.
The count field of $T[h_i(x)]$ is also incremented as $T[h_i(x)].count=T[h_i(x)].count +1$ at the same time.
The operation {\sf Delete}$(x,y)$ removes the key-value pair $(x,y)$ from the IBLT. The process is the same as that of
{\sf Insert}$(x,y)$ except for decrementing the counter. The operation  {\sf Get}$(x)$ retrieves the value corresponding to 
the key $x$. This operation is realized as follows. If there exists $i \in [1,k]$ satisfying $T[h_i(x)].Count=1$, then {\sf Get}$(x)$ returns $T[h_i(x)].ValueSum$. Otherwise, {\sf Get}$(x)$ declares the failure of the operation. 

The last operation {\sf ListEntries}$()$ outputs all the key-value pairs in the IBLT by sequentially removing 
the entries with the counter value equal to one from the table. The details of the process is as follows.
We first look for $i \in [1,m]$ satisfying $T[i].Count=1$.  If there exists $i^*$ satisfying the condition $T[i^*].Count=1$, 
the key-value pair $(T[i^*].KeySum, T[i^*].ValueSum)$ is registered into the output list 
and then {\sf Delete}$(T[i^*].KeySum, T[i^*].ValueSum)$ is executed. This process is iterated until 
no cell with the counter value equal to one can be found.  It should be remarked that, in some cases,  {\sf ListEntries}$()$
fails to list all the entry in the IBLT. This is because a non-empty IBLT can have counter values larger than one
for $i \in [1,m]$. This failure event is called a {listing failure}. It is desirable that an IBLT is designed to decrease the frequency 
of the listing failure events as small as possible.

\subsection{Probabilistic Model}

It is clear that the probability of the listing failure event, which is called the {\em listing failure probability}, depends on the
definition of the probabilistic model for keys and  hash functions.
In this paper (except for Section \ref{lowering}), we assume the following model for keys and hash functions.
The hash functions $h_1,\ldots, h_k$ have domain $\{0,1\}^b$
and the key of the entries to be stored are independent random variables uniformly distributed over $\{0,1\}^b$.
The number of entries are assumed to be $n$.
The hash functions are assumed to be uniform such that 
$h_i(x)$ distributes uniformly in the range of $h_i$ when $x \in \{0,1\}^b$ obeys the uniform distribution.
The $m$-cells are split into $k$-subtables each of size $m/k$ and 
each hash function uniformly selects a cell in a subtable. In other words, the range of $h_i$  is $[(i-1)*(m/k)+1, i*(m/k)]$.

\section{Upper Bound on Listing Failure Probability}
\label{upper}

In this section, we will derive an upper bound on the listing failure probability.
The listing failure event occurs when a stopping set \cite{di}, which is a combinatorial 
substructure of a matrix, appears.  In order to evaluate the listing failure probability,
we need to enumerate the number of {\em stopping matrices} of given size.
A stopping matrix is a matrix with no row of weight one corresponding to the case where
no cells with counter value equal to one exists.

\subsection{Enumeration of Stopping Matrix}

The state matrix $B$ of an IBLT  can be represented by an $m \times n$ binary matrix where $m=\ell k$.
A row of the matrix $B$ corresponds to a cell and a column corresponds to an entry. The matrix $B$ 
can be divided into disjoint $k$-blocks with size $\ell \times n$. If the $(s,t)$-element of the $u$-th block of $B$ is one,
this means that the $t$-th entry is hashed to the $s$-th cell by using the $u$-th hash function.
Suppose that a sub-matrix $M'$ consisting of several columns of $M$ have no rows of weight one.
In such a case, {\sf ListEntries}$()$ fails to list all the entry in this table because $M'$ cannot be resolved in the peeling process.
If a binary matrix $M'$ does not have a row with weight one, 
$M'$ is said to be a stopping matrix. The existence of a stopping matrix in $B$ is the necessary and 
sufficient condition for the failure of a peeling process \cite{di}\cite{richardson}.

In our case, the state matrix $B$ is divided into $k$-subblocks corresponding to subtables.
It might be reasonable to consider a stopping matrix in a subblock
before discussing the probability of the event that $B$ includes a stopping matrix.

Let $S^{(\ell, n)}$ be the set of $\ell \times n$ binary matrices with column weight one; i.e., 
\[
    S^{(\ell, n)} \hspace{-0.5mm}\defeq \hspace{-0.5mm}\{(m_1,\ldots,m_n) \hspace{-0.5mm}\in \hspace{-0.5mm}\{0,1\}^{\ell \times n} \hspace{-0.5mm}\mid \hspace{-0.5mm} wt(m_i)=1, i \in [1,n] \},
\]
where $wt(\cdot)$ represents the Hamming weight function.
From this definition, it is evident that the cardinality of $S^{(\ell, n)}$ is $\ell^n$.
The number of the stopping matrices in $S^{(\ell, n)}$ is denoted by $z(\ell, n)$, which can be written as
\begin{equation}
z(\ell, n) \defeq \# \{M \in S^{(\ell,n)} \mid M \mbox{ is a  stopping matrix}\}.
\end{equation}
For convention, $z(0,0)$ is defined to be 1.

The next recursive formula  plays a key role to enumerate $z(\ell, n)$ which is required for 
evaluating an upper bound for the listing failure probability.
\begin{theorem}[Recursive formula on $z(\ell, n)$]
The following recursive relation 
\begin{equation}\label{recursive}
z(\ell, n) = \ell^n - \sum_{c=1}^{\min(\ell,n)} c! {\ell \choose c} {n \choose c} z(l-c,n-c)
\end{equation}
holds for $\ell \ge 1$ and $n \ge 1$.
\end{theorem}
(Proof)
Let $a(\ell,n)$ be the cardinality of non-stopping matrices $a(\ell, n) \defeq \ell^n - z(\ell,n)$.
In the following, we enumerate $a(\ell,n)$ by using a recursive relation.
For given $M \in S^{(\ell, n)}$,
a pair $(i,j) \in [1,\ell] \times [1,n]$ is said to be a {\em pivot} of $M$ if
$M_{i,j} = 1$ and the Hamming weight of the $i$-th row of $M$ is 1.
The set of pivots of $M$ is denoted by
\[
piv(M) \defeq \{(i,j) \in [1,\ell] \times [1,n] \mid (i,j) \mbox{ is a pivot of } M\}.
\]
Note that $M$ is a stopping matrix if and only if $piv(M)$ is empty.
The cardinality of non-stopping matrices $a(\ell, n)$ can be represented by
\begin{equation} \label{aln}
a(\ell, n) = \sum_{i=1}^{\min(\ell,n)} \#T_i^{(\ell, n)} 
\end{equation}
where 
\begin{equation}
T_i^{(\ell, n)} \defeq  \{M \in S^{(\ell,n)} \mid \# piv(M) = i \}, \; i \in [0, \min(\ell, n)].
\end{equation}
This is because the set of non-stopping matrices can be partitioned 
into disjoint sets $T_i^{(\ell, n)}$ for $i \in [1, \min(\ell,n)]$.
In the following,  we will try to prove the equality
\begin{equation} \label{Ti}
\#T_i^{(\ell, n)}  = c! {\ell \choose c} {n \choose c} z(l-c,n-c)
\end{equation}
for $i \in [1, \min(\ell,n)]$.
Assume that $M \in T_c^{(\ell, n)}$ is given ($c \in [1, \min(\ell,n)]$).
By getting rid of all the column and rows corresponding to $piv(M)$ from $M$, 
we obtain an $(\ell-c) \times (n-c)$ matrix $M'$.
Namely we delete the $i$-th row and the $j$-th column from $M$ if
$(i,j) \in piv(M)$. From the assumption $M \in T_c^{(\ell, n)}$, the resulting 
matrix $M'$ must be a stopping matrix in $T_0^{(\ell-c, n-c)}$.
Note that the size of $T_0^{(\ell-c, n-c)}$ is given by $z(\ell-c, n-c)$.
Therefore, the size of $T_i^{(\ell, n)}$ is the product of the number of possible ways to 
choose $piv(M)$ and $z(\ell-c, n-c)$. Based on a simple combinatorial argument, 
we can see that the number of possible ways to choose $piv(M)$ can be enumerated as $c ! {\ell \choose c} {n \choose c}$.
As a result, we have the equality (\ref{Ti}). Combining (\ref{aln}) and (\ref{Ti}), 
the claim of the theorem is obtained. \hfill\qed

For some special combinations of $\ell$ and $n$,
$z(\ell, n)$ has a simple expression as follows.
\begin{eqnarray} \label{boundary1}
z(\ell, 1) &=& 0,\quad \ell \ge 1  \\
z(\ell, 2) &=& \ell,\quad \ell \ge 1  \\
z(\ell, 3) &=& \ell,\quad \ell \ge 1  \\ \label{boundary2}
z(1, n) &=& 1,\quad n \ge 1. 
\end{eqnarray}
These expressions can be easily proved based on the definition of the stopping matrix and of $S^{(\ell,n)}$.
The recursive formula (\ref{recursive}) enable us to evaluate the value of $z(\ell, n)$ efficiently.
These simple expressions can be used as boundary conditions for a recursive evaluation process.

Table \ref{tblzln} presents the values of $z(\ell,n)$ for $(\ell,n) \in [1,10]^2$. These values are 
computed based on the recursive formula (\ref{recursive}). Note that $S^{(\ell,n)}$ contains 
$10^{10}$-matrices when $\ell = n = 10$. A naive enumeration scheme generating all the matrices in $S^{(\ell,n)}$
may have computational difficulty even for such small parameters.

%

\begin{table}
    \caption{Values of $z(\ell, n)$:  Number of Stopping Matrices in $S^{(\ell,n)}$}
    \label{tblzln}
    {\tabcolsep=1.05mm
    \begin{tabular}{c|cccccccccc}\hline\hline
        $\ell \: \backslash \, n$ & 1 & 2 & 3 & 4 & 5 & 6 & 7 & 8 & 9 & 10\\ \hline
        1      & 0 & 1 & 1 & 1 & 1 & 1 & 1 & 1 & 1 & 1\\ 
        2      & 0 & 2 & 2 & 8 & 22 & 52 & 114 & 240 & 494 & 1004\\ 
        3      & 0 & 3 & 3 & 21 & 63 & 243 & 969 & 3657 & 12987 & 43959\\ 
        4      & 0 & 4 & 4 & 40 & 124 & 664 & 3196 & 15712 & 79228 & 396616\\ 
        5      & 0 & 5 & 5 & 65 & 205 & 1405 & 7425 & 44385 & 271205 & 1666925\\ 
        6      & 0 & 6 & 6 & 96 & 306 & 2556 & 14286 & 100176 & 691146 & 4916436\\ 
        7      & 0 & 7 & 7 & 133 & 427 & 4207 & 24409 & 196105 & 1471519 & 11773699\\ 
        8      & 0 & 8 & 8 & 176 & 568 & 6448 & 38424 & 347712 & 2775032 & 24547664\\ 
        9      & 0 & 9 & 9 & 225 & 729 & 9369 & 56961 & 573057 & 4794633 & 46341081\\ 
        10     & 0 & 10 & 10 & 280 & 910 & 13060 & 80650 & 892720 & 7753510 & 81163900\\ \hline
    \end{tabular}
    }
\end{table}

\subsection{Listing Failure Probability and its Bound}

The set of all the state matrix is defined as 
\begin{equation}
B^{(\ell, n, k)} \defeq \{(M_1,\ldots, M_k)^T \mid M_i \in S^{(\ell,n)}, i \in [1,k] \}. 
\end{equation}
The cardinality of $B^{(\ell, n, k)}$ is $\ell^{nk}$.
According to the scenario we have discussed in the previous section,  we here define a probability space by 
assigning the equal probability $1/\ell^{nk}$ to each element in $B^{(\ell, n, k)}$.

Suppose that $P_f(\ell,n,k)$ represents the listing failure probability, which 
is the probability that {\sf ListEntries}$()$ operation fails to list all the entries in the IBLT.
The next theorem provides an upper bound on $P_f(\ell,n,k)$.

\begin{theorem}[Upper bound on listing failure probability]
For given $\ell\geq 1,n\geq 1,k\geq 1$,  the listing failure probability $P_f(\ell,n,k)$ can be upper bounded by 
\begin{equation}
P_f(\ell,n,k) \le \sum_{i=2}^n {n \choose i} \left(\frac{z(\ell,i) }{\ell^i} \right)^k.
\end{equation}
\end{theorem}
(Proof)
The peeling process of the {\sf ListEntries}$()$ fails to recover all the entries in 
the IBLT if and only if $B \in B^{(\ell, n, k)}$ contains a stopping matrix as its sub-matrix.
Thus, $P_f(\ell,n,k)$ can be characterized as 
\begin{equation}
P_f(\ell,n,k) = Pr[B \mbox{ includes a stopping matrix}].
\end{equation}
For an index set ${\cal I} \in 2^{[1,n]}$, let $B_{\cal I}$ be the sub-matrix of $B$ 
consisting of columns of $B$ with indices in ${\cal I}$.
If $B_{\cal I} \mbox{ is a stopping matrix}$, then the index set ${\cal I}$ is said to be a {\em stopping set}.
The probability $P_f(\ell,n,k)$ can be upper bounded as follows:
\begin{eqnarray} \nonumber
P_f(\ell,n,k) &=& Pr[B \mbox{ includes a stopping matrix}] \\
&=& Pr\left[ \bigcup_{{\cal I} \in 2^{[1,n]} \backslash \emptyset }B_{\cal I} \mbox{ is a stopping matrix} \right] \nonumber \\
&\le& \sum_{{\cal I} \in 2^{[1,n]} \backslash \emptyset } Pr\left[ {\cal I} \mbox{ is a stopping set} \right].
\end{eqnarray}
The last inequality is due to the union bound.  From the definition of the probability space defined on $B^{(\ell, n, k)}$, 
the probability that ${\cal I}$ is a stopping set is given by
\begin{equation}
Pr\left[ {\cal I} \mbox{ is a stopping set} \right] = \left(\frac{z(\ell,\#{\cal I}) }{\ell^{\#{\cal I} }} \right)^k.
\end{equation}
By using this equality, we have the following upper bound:
\begin{eqnarray} \nonumber
P_f(\ell,n,k) 
\hspace{-2mm} &\le&\hspace{-2mm} \sum_{{\cal I} \in 2^{[1,n]} \backslash \emptyset } Pr\left[ {\cal I} \mbox{ is a stopping set} \right] \\ \nonumber
\hspace{-2mm}& =&\hspace{-2mm} \sum_{i=1}^n \sum_{ {\cal I} \in 2^{[1,n]} \backslash \emptyset} Pr\left[ {\cal I} \mbox{ is a stopping set} \hspace{-0.5mm}\mid \hspace{-0.5mm} \#{\cal I} = i \right] \\
\hspace{-2mm} &=& \hspace{-2mm}\sum_{i=2}^n {n \choose i} \left(\frac{z(\ell,i) }{\ell^i} \right)^k.
\end{eqnarray}
In the last equality, we used the fact $z(\ell, 1) = 0$. \hfill\qed

\section{Computer Experiments}
\label{numerical}

In this section,  we will present several results on computer experiments and on numerical 
evaluation of the upper bound presented in the previous section.

\begin{figure}[hptb]
  \begin{center}
   \includegraphics[width=88mm]{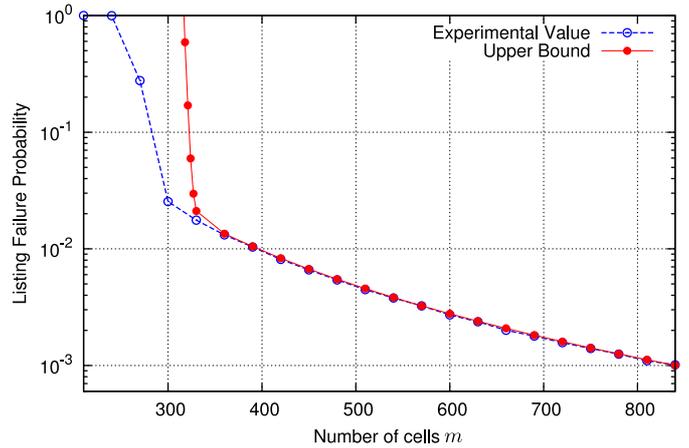}
  \end{center}
  \caption{Comparison of the listing failure probability: experimental values and upper bound ($n=210, \; k=3, \; b=32$).}
  \label{fig:210-30-840}
\end{figure}

In order to examine the tightness of the bound, 
Figure \ref{fig:210-30-840} presents 
curves of the listing failure probability obtained by computer experiments (dashed line) 
and of the upper bound (solid line).
These curves are plotted as functions of the number of cells $m$.
The number of entries is $n=210$ and the symbol size of the key is $b=32$.
In computer experiments, the number of trials is $10^6$. As a hash function, SHA-1\cite{SHA} was used.
The number of the hash functions assumed to be $k=3$.
We used pseudorandom $32$-bit numbers for pseudorandom key-value pairs.
It can be observed that the upper bound gives fairly tight estimation, as the number of cells $m$ increases.
As in the case of LDPC codes,  the error curve in Figure \ref{fig:210-30-840} exhibit both water fall and error floor phenomenon.
This result indicates that the upper bound precisely captures the error floor behavior of the listing failure probability.

\begin{figure}
  \begin{center}
   \includegraphics[width=88mm]{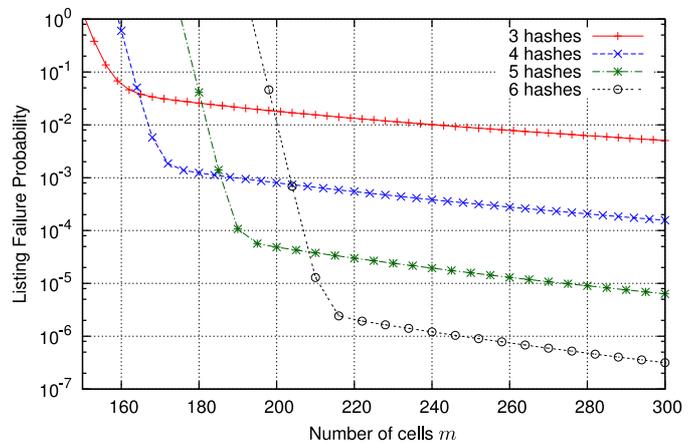}
  \end{center}
  \caption{Comparison of the upper bound on listing failure probability: 3 hashes, 4 hashes, 5 hashes and 6 hashes ($n=100$).}
  \label{fig:comp-hashes-org}
\end{figure}

From the upper bound, it is possible to see a tradeoff between the water fall and error floor.
Figure \ref{fig:comp-hashes-org} presents the upper bounds for $k \in [3,6]$.
The number of entries is $n=100$. A curve of the upper bound is plotted as a function of the number of cells $m$.
We can observe that the listing failure probabilities in the error floor region can be decreased as the number of hash functions 
$k$ increases. On the other hand,  increments of $k$ pushes the water falls to the right.

\begin{figure}[hptb]
  \begin{center}
   \includegraphics[width=88mm]{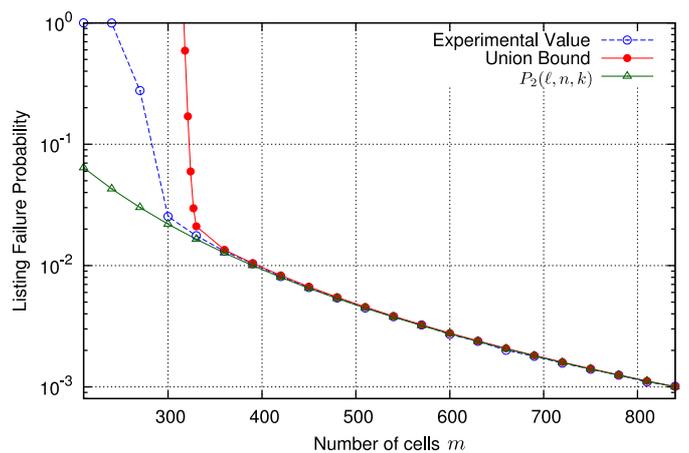}
  \end{center}
  \caption{Comparison of the listing failure probability: experimental values, upper bound and asymptote $P_{2}(\ell,n,k)$ ($n=210, \; k=3, \; b=32$).}
  \label{fig:SS-dist}
\end{figure}

From the upper bound and some experimental results, 
we see that stopping sets of size 2 dominates the error floor behavior.
Figure \ref{fig:SS-dist}
 presents the upper bound, the asymptote $P_2(\ell,n,k)$ defined by
\begin{equation}
    P_2(\ell,n,k)  \defeq {n \choose 2} \left(\frac{z(\ell,2) }{\ell^2} \right)^k = {n \choose 2} \frac{1}{\ell^k}
\end{equation}
and the experimental value of the list error probability. The result suggest that the probability of occurrence of stopping set of size 2 
determines the depth of an error floor.

\section{SS Avoiding Hash Function}
\label{lowering}

We have seen that stopping sets of size 2 dominate the behavior of the list 
failure probability in the error floor region. The stopping sets of size 2 occur 
when $k$-hash values for 2-distinct keys collide; i.e., 
\begin{equation}\label{collision}
(h_1(a), h_2(a),\ldots, h_k(a)) = (h_1(b), h_2(b),\ldots, h_k(b))
\end{equation}
for $a \ne b$. If this type of collision can be prevented,
it is expected that the error floor performance can be improved.

The SS avoiding  hash function defined here are designed so that the collisions (\ref{collision}) are avoided.
In the following discussion, we will further assume the uniqueness of keys registered in the IBLT.
Namely, an insertion of district entries with the same keys and a multiple insertion of the same key-value pairs
are not allowed.  This assumption may be natural for most of applications such as set reconciliation.

Let a hash function $h$ be an bijective map from $\{0,1\}^b$ to $\{0,1\}^{sk}$ where $b = sk$.
The {\em SS avoiding hash functions} $(h_1,\ldots, h_k)$ are simply defined by partitioning the output $sk$-tuple from $h$
into $k$ binary $s$-tuples; i.e., $h_i(x)$ is given by
\begin{equation}
h_i(x) = q_i + (i-1) 2^s+1,\quad i \in [1,k],
\end{equation}
where $(q_1,\ldots, q_k)=h(x) (q_i \in \{0,1\}^s)$.
Note that $m/k = 2^s$ holds; i.e., each subtable contains $2^s$-cells.
Due to the assumption on the uniqueness of the keys in the IBLT,
it is evident that a collision (\ref{collision}) does not occur. This means that 
occurrences of the stopping sets of size 2 can be completely prevented.
Note that the use of the SS avoiding hash function introduces a restriction on several system parameters; i.e., $b = sk$.
This inflexibility can be considered as a price to be paid for lowering the error floor.

It should be remarked that the probabilistic model assumed in Section \ref{pre} cannot 
be directly applied to the system presented in this section
This is because the assumption on the uniqueness of the keys introduces weak correlations 
between the stored entries. Although we have to take care of these distinctions, the analysis presented in the previous sections 
may be still useful for predicting the performance of {\sf ListEntries}$()$ with the SS avoiding hash functions 
if $b$ is large enough.

\begin{figure}[b]
  \begin{center}
   \includegraphics[width=88mm]{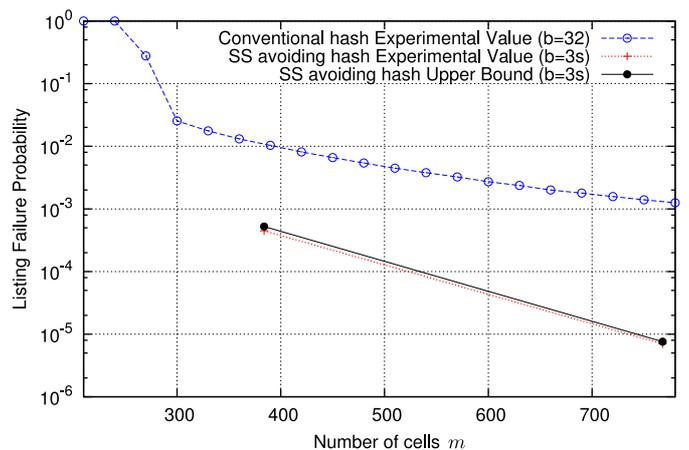}
  \end{center}
  \caption{Comparison of the listing failure probability: conventional hash function and SS avoiding hash function ($n=210, \; k=3$).}
  \label{fig:SSavoid}
\end{figure}

Figure \ref{fig:SSavoid} presents the results of a computer experiment on 
the SS avoiding hash functions. As an bijective map, the identity map was exploited.
The two curves of listing failure probabilities are plotted;
the first one corresponds to the case of a conventional hash function and the second one 
corresponds to the case of the SS avoiding hash function where the symbol size of the key is $b=3s$.
In both cases, the number of entries is $n = 210$ and  the number of hash functions is assumed to be $k=3$. 
We can observe that the SS avoiding hash function reduces the listing failure probabilities in the error floor region.
Furthermore, the upper bound almost captures the error floor behavior of the listing failure probability in this settings.

\section{Conclusion}
\label{conclusion}
In this paper, we presented a finite length performance analysis on the listing failure probability which 
may be useful for designing a system or an algorithm  including the IBLT as a building component.  
The recursive formula presented in Section \ref{upper} will become an useful tool for finite length analysis.
In Section \ref{numerical}, we have seen that the error floor performance can be improved by increasing the number 
of the hash functions but it degrades the waterfall performance.
From the results shown in Section \ref{lowering}, 
we can expect that
appropriately designed SS avoiding hash functions can improve the error floor performance without sacrificing the waterfall 
performance.


%



\begin{thebibliography}{99}
    \bibitem{BF}B. Bloom,
        ``Space/time trade-offs in hash coding with allowable errors,'' {\it Communications of the ACM}, vol. 13, no. 7, pp. 422-426, 1970.
    \bibitem{bonomi1}F. Bonomi, M. Mitzenmacher, R. Panigrahy, S. Singh, and G. Varghese,
        ``Beyond Bloom filters: From approximate membership checks to approximate state machines,'' {\it ACM SIGCOMM Computer Communication Review}, vol. 36, no. 4, pp. 326, 2006

    \bibitem{bonomi2}F. Bonomi, M. Mitzenmacher, R. Panigrahy, S. Singh, and G. Varghese,
        ``An improved construction for counting Bloom filters,'' {\it In Proceedings of the
        European Symposium on Algorithms (ESA)}, vol. 4168 of {\it LNCS}, pp. 684-695, 2006.

    \bibitem{broder}A. Broder and M. Mitzenmacher,
        ``Network applications of Bloom filters: A
        survey,'' {\it Internet Mathematics}, vol. 1, no. 4, pp. 485-509, 2004.

    \bibitem{chazelle}B. Chazelle, J. Kilian, R. Rubinfeld, and A. Tal,
        ``The Bloomier filter: an efficient data structure for static support lookup tables,'' 
        {\it In Proceedings of the Fifteenth Annual ACM-SIAM Symposium on Discrete Algorithms}, pp. 30-39, 2004.

    \bibitem{CBF}M. Mitzenmacher,
    ``Compressed Bloom filters,'' {\it IEEE/ACM Transactions on Networking},
        vol. 10, no. 5, pp. 613-620, 2002.

    \bibitem{IBLT} M. Goodrich and M. Mitzenmacher,
        ``Invertible bloom lookup tables,'' {\it In Proceedings of the 49th Allerton Conference}, pp. 792-799, 2011.

    \bibitem{putze}F. Putze, P. Sanders, and J. Singler,
        ``Cache-, hash-, and space-efficient Bloom filters,'' {\it ACM Journal of Experimental Algorithms}, vol. 14, pp. 4.4-4.18, 2009.

    \bibitem{eppstein}D. Eppstein and M. T. Goodrich,
    `` Straggler identification in round-trip data streams via Newton's identities and invertible Bloom filters,'' {\it IEEE Transactions on Knowledge and Data Engineering}, vol. 23, no. 2, pp. 297-306, 2011.

    \bibitem{BC}M. Mitzenmacher, G. Varghese,
        ``Biff (Bloom filter) codes: fast error correction for large data sets,''
        {\it Information Theory Proceedings (ISIT), IEEE International Symposium on}, pp. 483-487, 2012.

    \bibitem{di}C. Di , D. Proietti , I.E.Teletar, T. Richardson and R. Urbanke,
        ``Finite-length analysis of low-density parity-check codes on the binary erasure channel,''
        {\it IEEE Transactions on Information Theory}, vol. 48, no. 6, pp. 1570-1579, 2002.

        \bibitem{richardson}T. Richardson and R. Urbanke, {\it Modern Coding Theory}, Cambridge University Press, 2008.

    \bibitem{SHA}National Institute of Standards and Technologies, {\it Secure Hash Standard}, Federal
    Information Processing Standards Publication, FIPS-180, 1993.
\end{thebibliography}
\end{document}